# Predicting Human Brain States with Transformer

Yifei Sun[1,2], Mariano Cabezas[2,3,4], Jiah Lee[1], Chenyu Wang[2,3,5], Wei Zhang[6],

Fernando Calamante[1,2,7], Jinglei Lv[1,2]

[1] School of Biomedical Engineering, University of Sydney, NSW 2008, Australia.
[2] Brain and Mind Center, University of Sydney, NSW 2050, Australia.
[3] Central Clinical School, University of Sydney, NSW 2050, Australia.
[4] Macquarie University Hearing, Macquarie University, NSW 2109, Australia.
[5] Sydney Neuroimaging Analysis Centre, University of Sydney, NSW 2050, Australia.
[6] School of Computer and Cyber Sciences, Augusta University, GA 30901, United States.
[7] Sydney Imaging, University of Sydney, NSW 2006, Australia.
`yifei.sun@sydney.edu.au`

**Abstract.** The human brain is a complex and highly dynamic system, and our current knowledge of its functional mechanism is still very limited. Fortunately, with functional magnetic resonance imaging (fMRI), we can observe blood oxygen level-dependent (BOLD) changes, reflecting neural activity, to infer brain states and dynamics. In this paper, we ask the question of whether the brain states represented by the regional brain fMRI can be predicted. Due to the success of self-attention and the transformer architecture in sequential auto-regression problems (e.g., language modelling or music generation), we explore the possibility of the use of transformers to predict human brain resting states based on the large-scale high-quality fMRI data from the human connectome project (HCP). Current results have shown that our model can accurately predict the brain states up to 5.04s with the previous 21.6s. Furthermore, even though the prediction error accumulates for the prediction of a longer time period, the generated fMRI brain states reflect the architecture of functional connectome. These promising initial results demonstrate the possibility of developing generative models for fMRI data using self-attention that learns the functional organization of the human brain. Our code is available at: https://github.com/syf0122/brain_state_pred

## 1 Introduction

The human brain is an intricate dynamic system with tens of billions of neurons and trillions of synaptic connections [1]. Understanding the dynamic mechanisms of the human brain [2] is always a top priority of neuroscience, as it is essential for uncovering the origins of cognition, emotion, language and other higher-level human intelligence [3]. Additionally, this understanding is crucial for deciphering the mechanisms behind brain disorders such as Alzheimer's Disease [4, 5] and Schizophrenia [6, 7]. Moreover, as brain-computer interfaces (BCI) and brain-inspired artificial intelligence become the current technology trend, learning the mechanisms for brain dynamics [8, 9] becomes an essential step to mimicking them.

Functional magnetic resonance imaging (fMRI) is a widely used non-invasive and in-vivo technique to observe whole brain dynamics spatially at the meso-scale and temporally at the second scale [10]. Despite significant progression in mapping the brain's functional organization [11] (e.g., the intrinsic networks were reconstructed with resting-state fMRI [12]), understanding its functional connectivity (FC) has become an important biomarker for healthy brains and mental health research [4, 6]. While recent deep learning approaches to model the hierarchical organization of brain function have become critical achievements in the field [13, 14], the question of how brain activity emerges even when we are not performing a specific task (i.e., resting state) still remains unanswered [15]. Moreover, whether specific future sequential brain states from a resting state acquisition can be predicted still remains unknown. Addressing this technology and knowledge gap could potentially reduce the fMRI scan time for patients with difficulties or disabilities. If the brain states can be predicted, the pain and harm from certain fatal brain disorders, such as epilepsy, can also be avoided or at least reduced [16, 17]. Furthermore, the ability to predict brain states paves the way for BCI technologies, potentially enabling more intuitive and effective communication. This capability could transform not only medical therapeutics but also how to interact with technology, making artificial intelligence systems more adaptive to our cognitive and emotional states.

Since the introduction of multi-headed self-attention blocks by Vaswani et al. [18] in 2017, transformer architectures have become ubiquitous in deep learning for multiple tasks focusing on sequences [19, 20] and images [21, 22]. A recent success story is ChatGPT [23], which exemplifies the power of transformers in processing sequential information in natural language that learns patterns from a knowledge pool and gives answers within the context of a continuous conversation. Given their ability to find long-distance relationships between data tokens (in our case, brain states) grounded in correlations and with links to graph theory, we believe that self-attention-based architectures have the ability to learn patterns from sequential brain activity and predict the upcoming brain states. More recently, Peter et al. [24] and Malkiel et al. [25] demonstrated the potential of the transformer framework in analyzing fMRI data for age prediction, gender classification, and disease classification. Furthermore, a brain language model (BrainLM) [26] was proposed as a foundation model for brain dynamic activities, especially focusing on the fMRI data. The BrainLM is pre-trained for masked prediction and then fine-tuned for future brain state prediction, outperforming other state-of-the-art architectures. This further shows the use of self-attention-based frameworks is promising in brain state prediction. However, the BrainLM needs the use of a large dataset for pretraining and the brain states prediction requires a relatively long time series (180 time points). If we can train a model to predict future brain states depending on a shorter input time series, we can largely shorten the scanning time.

In this paper, we frame the brain state prediction problem as an auto-regression task where given a sequence, we predict the next temporal element. Specifically, we propose a novel method based on the time series transformer [27] architecture to predict future brain states comprising 379 grey matter regions of a whole brain given a sequence of previous time points as observed in an fMRI acquisition. We train our model on high-quality data from the Human Connectome Project (HCP) and present a set of promising results on brain state prediction. Briefly, the model can accurately predict the immediate brain state, predict the brain states of 5.04 seconds with low error, and predict brain states of over 10 minutes agreeing with the average human functional connectome.

## 2   Method

### 2.1   Data and Preprocessing

In this study, the resting-state fMRI (rs-fMRI) data from the HCP young adults dataset [28] were used. We employed the 3 Tesla fMRI data of 1003 healthy young adults, excluding 110 subjects with missing or incomplete rs-fMRI scans. For each subject, four rs-fMRI scans, with 1200 time points each in the CIFTI [29] format that stores surface-based gray matter data, were utilized. The HCP rs-fMRI data has an isotropic spatial resolution of 2mm and a temporal resolution of 0.72s.

Aside from the minimal preprocessing already provided with the HCP dataset [29], we carried out several additional preprocessing steps to further clean the data and prepare the data for training and testing the transformer. First, we spatially smoothed the fMRI data using a Gaussian filter with the full width at the half maximum (FWHM) set to 6mm in the CIFTI format, in order to reduce noise and improve signal-to-noise ratio [30]. Then, a bandpass filter was employed to filter out some uninterested noises while retaining the temporal signal within the 0.01 to 0.1 Hz range [31]. In order to bring all the samples onto a common scale, we applied the z-score transformation on time series to get zero temporal mean and unit standard deviation. Finally, the mean fMRI time series were calculated for 379 brain regions, including 360 cortical regions and 19 subcortical regions, using the multi-modal parcellation (MMP) atlas [32]. Therefore, we use a vector with the signal intensity of 379 regions at each time point to represent a brain state.

### 2.2   Transformer model

The human brain is a dynamic system, where its current state is related to the previous ones. In this paper, we explore the possibility of predicting a single brain state given a sequence of previous brain states. To model these predictions, we re-designed an existing time series transformer developed for influenza forecasting [18, 19, 27]. Our model comprises a combination of transformer encoders and decoders, with the whole architecture illustrated in Fig. 1.

The transformer takes as input time series data represented by a sequence of tokens with a given window size. As self-attention treats token relationships as a graph (ignoring token order), positional encoding with the sine and cosine functions is used to add relative temporal information [18, 27]. The encoder of the network contains four encoding layers with self-attention and feed-forward. These layers contain eight attention heads. Ultimately, this encoding stack generates the encoder output.

The last time point of the encoder input combined with the encoder output serves as the input of the decoder, which is defined as a stack of four decoding layers, which are also comprised of self-attention (with eight attention heads) and feed-forward layers. Finally, a fully connected layer maps the output from the decoder layers stack to the target output shape. Our model, unlike the time series transformer for influenza prevalence cases [27] that predicts a series of future time points and employs look-ahead masking to ensure that the predictions are based on past data [18], focuses on predicting a single future time point (*i+1*) using the immediate previous time point (*i*) and the encoder output. Therefore, we have omitted the use of look-ahead masking, simplifying the prediction process.

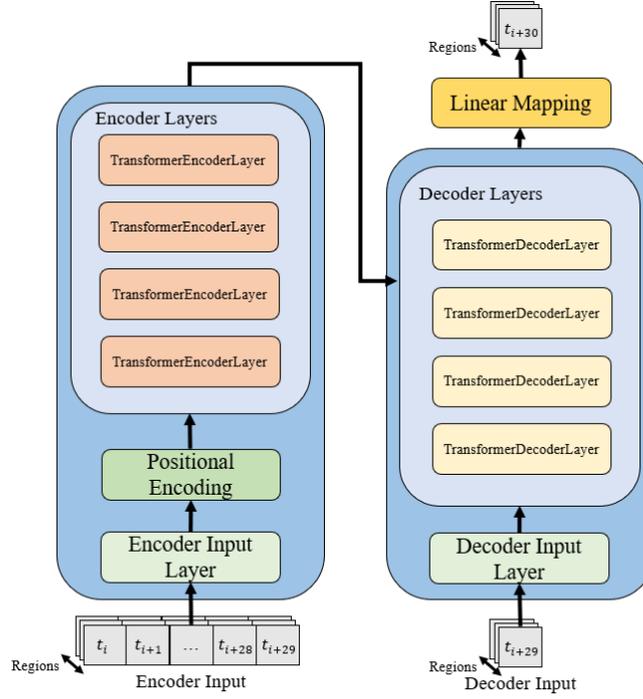

**Fig. 1.** Scheme of the transformer architecture for predicting brain states. The encoder takes an input of 30 ($t_i - t_{i+29}$) time points of 379 regions, the decoder takes the last input time point ($t_{i+29}$), and the encoder output to predict the next time point ($t_{i+30}$) of 379 regions.

### 2.3 Training

As we framed the brain state prediction problem as an autoregression task, we used the mean squared error (MSE) as the loss function and the Adam optimizer [33] with an initial learning rate of $10^{-4}$ to train our model.

We first conducted a primary test with different window sizes, i.e. the number of input timepoints, (5, 10, 15, 20, 30, 40, and 50) and the number of epochs per run (20, 25, 30, and 50) with forty subjects' data. All transformers started their prediction on the 51$^{st}$ time point to ensure a fair comparison. We allowed the overlap between input time series, so there are 1150 training samples for each rs-fMRI session data. During training, samples were randomly chosen from all training data across all subjects and all sessions, with a batch size of 512 to speed up the training. The model performance was then evaluated and compared using ten independent subjects not previously seen.

After the preliminary test, the best model was determined to be the one with a window size of 30, and after 20 epochs the model converged with the data of forty subjects. Thus, the window size was set to 30 for training the model with a much larger dataset. Additionally, ten-fold cross-validation was employed for model training and validation. In each fold, among the 1003 subjects, 90% of subjects (roughly 901 or 902 subjects) were used to train the model, while the remaining 10% of subjects (around 102 or 101 subjects) were used for validation. The number of training epochs was set to 10 because the training and validation losses were stable after the 6 epochs without overfitting.

### 2.4 Evaluation

After training the transformer network, we evaluated its performance with the rs-fMRI data of subjects that were not previously seen by the model. First, we tested the ability of the model to predict a single brain state from true fMRI data. Then, we performed a similar test with the same input sequences where the order of the brain states was randomized. We hypothesized that a model that truly learns sequential information and brain dynamics should produce a higher error when fed the same data in a randomized sequence. To test this hypothesis, we calculated the MSE for both tests for comparison and conducted a paired t-test on the two sets of MSE results.

After that, we evaluated the ability of our model to predict a series of brain states using limited true fMRI data and an increasing rate of synthesized states. Specifically, we used the thirty true fMRI time points to predict the next time point and then concatenated this prediction to the true time series and shifted the input window by one step to include the new predicted time point in an iterative manner until a sequence of the same length of time series as the true data (1200 time points) was synthesized. This resulted in the prediction of a synthetic time series of 1150 time points. MSE was calculated between the predicted time series and the true fMRI data together with Spearman's correlation between each predicted and true brain state to test for monotonic correlations.

Finally, the FC matrices for both true and predicted fMRI time series were calculated using Pearson's correlation between regional time series. Only the predicted portion of the brain states (1150 time points) was used for the FC analysis to avoid leakage from the true signal. To measure the similarity between the predicted FC matrices and the group average FC matrix based on the true fMRI data, we first calculated the mean absolute difference between the predicted FC matrix of each individual and the group average FC matrix calculated with true data. Spatial correlations between the predicted and true FC matrices were also calculated using the flattened upper triangle of FC matrices.

## 3    Results and Discussion

### 3.1    Model Selection

To select the proper window size and the number of training epochs, we evaluated the error of the trained models in predicting single brain states at one time point and the whole time series prediction using a subset of the dataset as a preliminary experiment.

As shown in Fig. 2a, when the model was trained for twenty epochs, it predicted single brain states with the least MSE, and using window sizes of fifteen and thirty showed a lower MSE. By looking at the MSE of the whole time series of generated brain states (Fig. 2b), models with window sizes of twenty and thirty showed the best performance. Thus, for the rest of the experiments, we used the model trained with a window size of thirty, but the number of epochs was set based on the observation of the actual training and validation losses of the model on the entire dataset because there are more training samples.

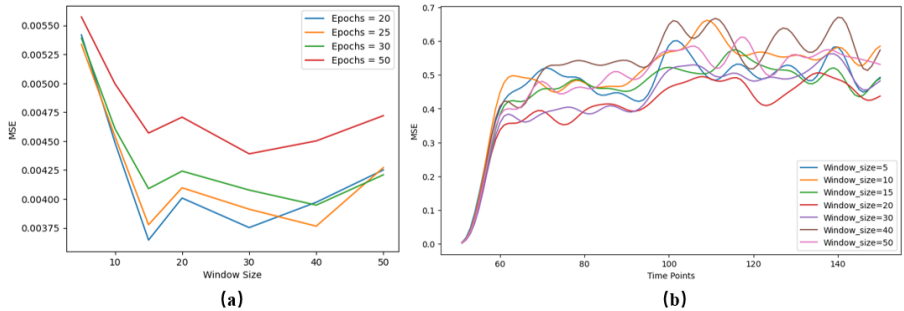

**Fig. 2.** Performance of the model using different window sizes. (a) The MSE of the models trained with different numbers of epochs and window sizes when predicting a single time point using true fMRI data. (b) The MSE of the models trained with 20 epochs and different window sizes when predicting long time series using the previous predicted time points (predicted brain states are used for future prediction).

### 3.2    Single Next Time Point Prediction

When we evaluated the model on the true fMRI data for single next time point predictions, the average MSE for all time points was 0.0013. However, when we shuffled the time points in the input time sequence, the MSE increased to 0.97 (more than 700 times higher, p-value $<10^{-10}$). These results provide strong evidence that the transformer model effectively learned and leveraged the temporal dependencies present in the fMRI time series. The substantial increase in error upon shuffling indicates that the brain state prediction accuracy is indeed driven by its ability to understand the sequential nature of the data, which is crucial for accurately modelling brain activity patterns over time. Thus, this shows the transformer's potential for applications in studying the functional dynamics of the human brain.

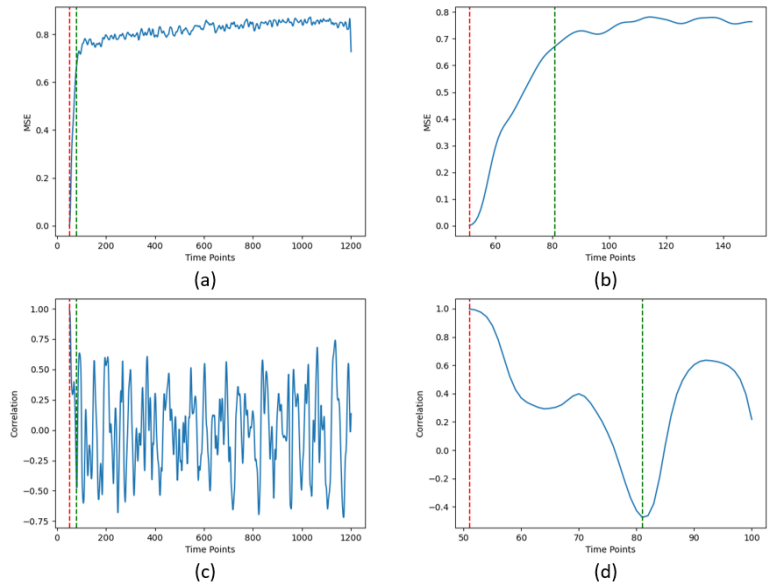

**Fig. 3.** Time series prediction MSE and correlation between the predicted and true brain states. The red dashed line specifies the first prediction starting point, which is the MSE of the 51st time points. The green dash line shows the point where the time points after it (on the right-hand side) are predicted purely depending on previously predicted data. Time points between two dashed lines are predicted partially based on real data. (a) The MSE of all predicted time points averaged across all test subjects. (b) The zoomed-in version of (a), showing the MSE of the first 100 predicted time points averaged across all test subjects. (c) The correlation coefficients of all 1150 predicted brain states averaged across all test subjects. (d) The zoomed-in version of (c), showing the correlation coefficients of the first 50 predicted brain states averaged across all test subjects. The red and green dash lines possess correspondence across the 4 sub-figures.

### 3.3 Time Series Prediction

As shown in Fig. 3b, the prediction MSE rises after predicting several sequential time points (after ~20 timepoints). We hypothesized that this is caused by cascading errors that accumulate as we sequentially predict brain states with more synthetic states as input. The accumulation of prediction errors can be understood similar as a Markov chain [34], where each state is predicted based on previous states. As more previously predicted states are incorporated into subsequent predictions, the accumulated errors from prior steps propagate through the chain, leading to an increased overall prediction error. Fig. 3a shows that after the input to the transformer comprises only predicted data, the error slightly increases and starts to fluctuate. More importantly, the first seven predicted time points (within the time of 5.04s) have relatively low MSE ($< 0.15$) and the MSE for the first 20 time points is roughly 0.26. After that, the MSE increases to around the level of 0.80. The similar phenomenon is observed in BrainLM [26], where the prediction MSE rises gradually after the prediction of the first 20 time points. In comparison, their 20-time-point MSE is 0.568, larger than ours. Overall, our results of the time series prediction show that future brain states of around 5.04s can be accurately predicted using a short fMRI sequence of only 21.6s.

We also calculated Spearman's correlation between the predicted brain states and the true fMRI data for all predictions. According to Fig. 3c and 3d, the first five time points have high correlation coefficients >0.85 (especially the first time point with a correlation of 0.997 with the true data). After that, even though the correlation drops quickly and starts to fluctuate, it is evident that our method can accurately predict at least seven consecutive brain states (5.04s) based on 30 fMRI time points (21.6s).

### 3.4 Functional Connectivity

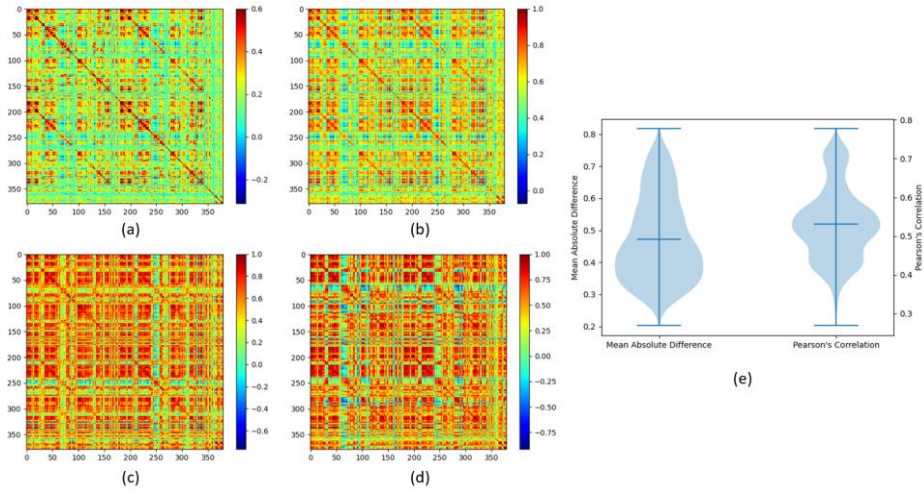

**Fig. 4.** FC matrices, mean absolute difference, and Pearson's correlation. (a) The group average FC calculated based on the true fMRI (b) The group average FC calculated based on predicted fMRI. (c) The FC calculated based on the predicted fMRI for one session of a subject. (d) The FC calculated based on the predicted fMRI for one session of another subject (e) Distribution of the mean absolute differences and Pearson's correlation coefficients between predicted and true group average FC. The mean absolute difference is shown on the left, and Pearson's correlation is shown on the right. The three horizontal lines specify the minimum, maximum, and mean of the mean absolute difference and Pearson's correlation, respectively.

FC matrices were calculated for both the true and predicted data. Fig. 4a and 4b show the group average FC matrices calculated from the HCP rs-fMRI data and the predicted time series for the same set of subjects. Although they have a different range, true and predicted FC show similar global patterns. Fig. 4c and 4d are the predicted FC of two individuals, that show similarity with the group average in Fig.4b, but clear individual variability.

The distribution of the mean absolute differences and spatial correlation between the predicted and true group average FC matrix is shown in Fig. 4e. From the figure plot, it is obvious that most of the predicted FC matrices have relatively low mean absolute differences and relatively strong correlations, with MSE from 0.35 to 0.45 and correlation from 0.50 to 0.60, respectively.

The relatively low mean absolute differences and relatively strong correlations confirm that the transformer model was able to capture the group-level functional organization. By predicting the group average connectivity patterns, this transformer captures the collective patterns shared by individuals in the HCP dataset, which can be beneficial for understanding the underlying principles of healthy brain function at the group level. In clinical cohorts, getting the group average FC allows the identification of common patterns that may serve as biomarkers of certain neurological conditions, aiding the development of diagnostic tools or treatment strategies.

## 4 Conclusion

In this work, we presented a novel method that employs a transformer network to predict brain states using fMRI data. The results demonstrated that our method could actually learn the temporal dependencies of brain states over time and accurately predict approximately 5.04s based on 21.6s of fMRI data. This shows the possibility of using a short fMRI segment for future brain state prediction. Furthermore, the model could learn the functional organization of the healthy human brain. In the future, we aim to improve this transformer architecture to generate more accurate predictions by mitigating the error accumulation issue. The improved prediction would be beneficial for studying the brain functions of vulnerable individuals who are not able to have their fMRI scanned for a long period. We also plan to develop the personalized model by using transfer learning. The explainability of our method will also be explored to understand the functional principles of the human brain.

## 5 Acknowledgements

J. Lv is supported by Brain and Mind Centre Research Development Grant, USYD-Fudan Brain and Intelligence Science Alliance Flagship Research Program, Moyira Elizabeth Vine Fund for Research into Schizophrenia Program and ARC Discovery Project (DP240102161).